# Communication Layer Security in Smart Farming: A Survey on Wireless Technologies

H. Mohammadi Rouzbahani, H. Karimipour, E. Fraser, A. Dehghantanha, E. Duncan, A. Green, C. Russell

*Abstract*—Human population growth has driven rising demand for food that has, in turn, imposed huge impacts on the environment. In an effort to reconcile our need to produce more sustenance while also protecting the world's ecosystems, farming is becoming more reliant on smart tools and communication technologies. Developing a smart farming framework allows farmers to make more efficient use of inputs thus protecting water quality and biodiversity habitat. Internet of Things (IoT), which has revolutionized every sphere of the economy, is being applied to agriculture by connecting on-farm devices and providing real-time monitoring of everything from environmental conditions to market signals, through to animal health data. However, utilizing IoT means farming networks are now vulnerable to malicious activities, mostly when wireless communications are highly employed. With that in mind, this research aims to investigate different utilized communication technologies in smart farming. Moreover, possible cyber-attacks are investigated to discover the vulnerabilities of communication technologies considering the most frequent cyber-attacks that have been happened.

*Keywords*— Smart Farming, Internet of Things (IoT), Communication Layer, Cyber-attack.

## I. Introduction

By the end of 2050, the world population will likely be 10 billion people, and at the same time, the food supply per capita increases ceaselessly [1]. A rising food demand, growing urbanization, climate change, and other global crises (e.g., COVID-19 pandemic) represent challenges that may lead to food insecurity [2].

Since conventional farming procedures cannot address these challenges, a modern framework is needed to maximize crop yield and profit while minimizing waste and environmental impact [3]. A fully interconnected multilayer architecture, also known as the IoT, is required to collect, transfer, and analyze data via different layers, including perception, communication, processing, and application layers [4]. Properly done, such a smart farming framework will give producers the ability to increase productivity while reducing environmental impact and waste through the precision application of inputs such as fertilizers and pesticides.

A smart farming framework is a multilayer system that consists of IoT infrastructures along with data processing capability. Several physical components, including different sensors, meters, surveillance cameras, Radio-frequency identification (RFID) tags, readers, and Global Positioning System (GPS) devices, collect required information as the perception layer [5]. The communication layer refers to a data transferring network between the perception and application layer employing short, medium, and long-range protocols [6]. Short-range networks like ZigBee and WiFi enable short-distance data transferring, while Lora is a protocol supporting middle-range communication [7]. Long-distance data transmission is provided through long-range protocols, including WiMAX, and cellular networks (e.g., GPRS, 3G, 4G, and 5G) [8]. In the processing layer, the collected data are stored before the cleaning, visualizing, and processing steps that are required to create a model. Finally, farmers access various management frameworks through the application layers in order to monitor the entire farm procedure. Fig. 1 shows a schematic of a smart farming framework.

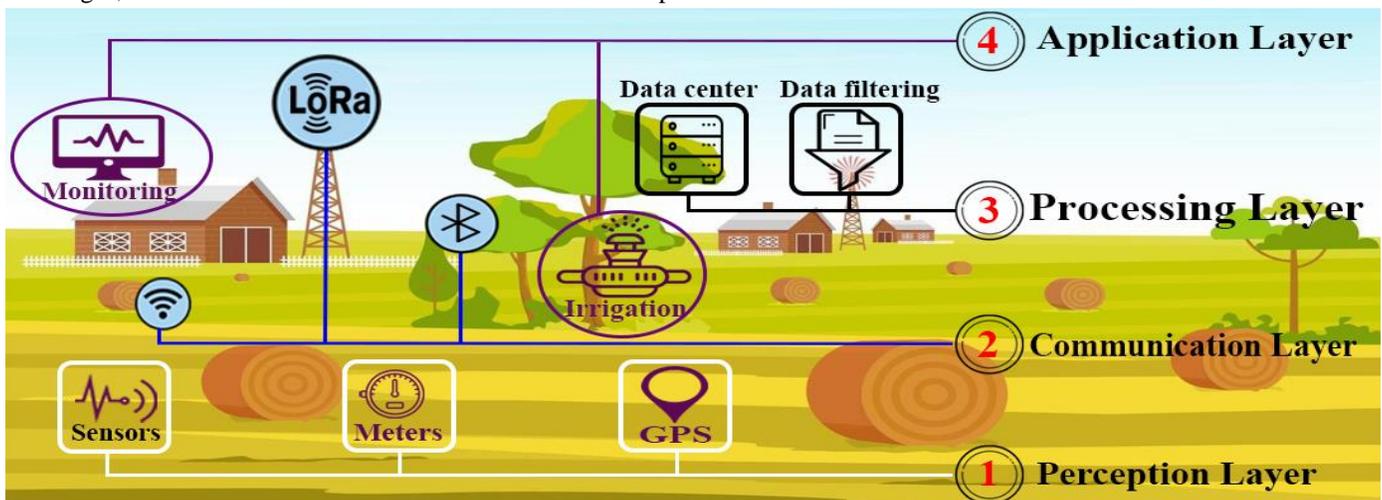

Fig. 1 Schematic of a smart farming framework

Hossein Mohammadi Rouzbahani is with University of Guelph, Canada (e-mail: hmoham15@uoguelph.ca).

Utilizing various sensors, smart meters, cameras, gateways, actuators, antenna, and transceivers makes an IoT-based farming framework more vulnerable to potential malicious activities and cyber-attacks [9]. Also, the communication layer is known as the most critical section from the security point of view since the common theme for smart frameworks, as well as smart farming, is communication and connectivity (from local to global scale). Moreover, considering the harsh, extensive, and highly distributed nature of farms, wireless data transmissions are the extremely employed type of communication that, as an open network environment, pave the way for attackers [10].

There have been several studies conducted on the security of the communication layer in smart farming. An attack prediction approach for smart farming networks has been presented by Jason West [11], focusing on false alarm reduction, while blockchain-based solutions have been suggested in [12] and [13]. In [14], a brief survey on security challenges in smart farming has been conducted, along with presenting a general attack taxonomy and studying different threat models.

Unauthorized access has been investigated by Rumyantsev et al. [15], offering a two-pass compensation based on fiber-optic communication, enhancing the efficiency by reducing misleading solutions. In [16], Support Vector Machine (SVM), K-Nearest Neighbors (KNN), and decision tree-based models have been employed in unauthorized access point datasets in order to find the best true positive rate, where the KNN and decision show satisfying performance. Effect of Denial of Service (DoS) attack on a smart farming structure's functionality has been investigated in [17], besides studying WiFi authentication attack caused by IEEE 802.11 vulnerabilities.

Intrusion attack is another security issue that has been studied in the literature. A two-level intrusion detection method has been developed in [18] against Modbus/Transmission Control Protocol (TCP) attack focusing on communication latency. In [19], a stereo depth intrusion detection algorithm has been established, reducing false positive and false negative rates simultaneously.

There are also many surveys on cybersecurity threats and challenges in IoT-based smart farming. In [20], the security challenges, basic characteristics, and performance of the application layer of smart farming frameworks were examined. Gupta et al. [21] investigated security and privacy issues in a multi-layered precision architecture as a highly distributed cyber-physical system. A comprehensive survey on security challenges in an IoT-based smart farming network has been presented by Demestichas et al. [22], focusing on associated emerging vulnerabilities.

None of the previous studies focused on the utilized protocols and associated security challenges in the communication layer of the smart farming network. This review investigates various communication technologies utilized in smart farming networks to determine their advantages and disadvantages by breaking it down into operation frequency band, protocols and standards, data rate, indoor and outdoor ranges. Moreover, potential cybersecurity issues and possible cyber-attacks are studied in order to discover the vulnerabilities and security deficiencies in the communication layer.

As Fig. 2 demonstrates, the communication types in an IoT-based smart farming platform are categorized into three main classes based on the coverage area radius difference, where the communication ranges are varied from less than one meter for Near-field communication (NFC) to more than 40 kilometers [23].

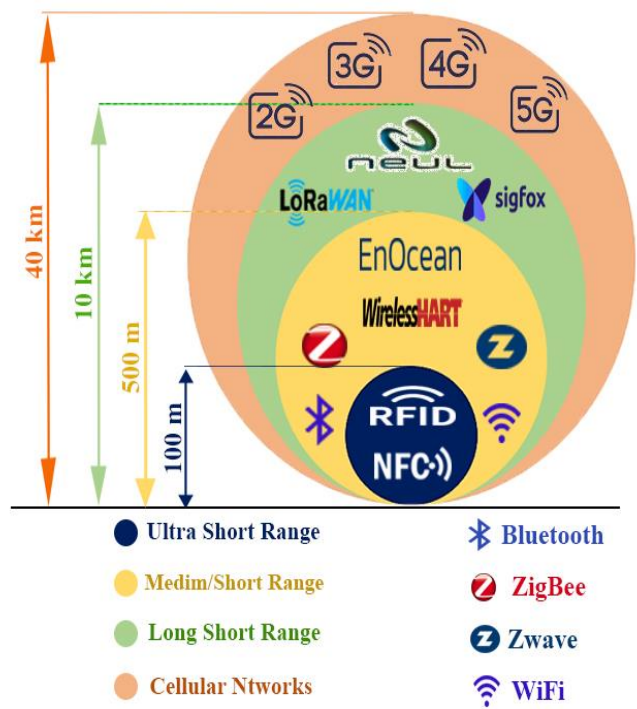

Fig. 2 Smart Farming Communication Technologies

The rest of this review paper is organized as follows: Section 2 provides an overview of communication technologies in smart farming. Section 3 introduces security challenges. Finally, section 4 presents conclusions.

II. COMMUNICATION TECHNOLOGIES

In this review paper, we categorize communication standards into four main classes: 1) ultra-short-range protocols, 2) short and medium range protocols, 3) long-range protocols, and 4) cellular networks. This categorization recognizes different parameters, including range, frequency band, data rates, power consumption, and ease of installation. These classifications are detailed below.

*A. Ultra-Short-Range Protocols*

Optimizing cost, energy consumption, and throughput are fundamental concerns at the first level of communication networks where numerous nodes are linked, and localization is prone to be a significant challenge. In order to address mentioned concerns, two primary standards have been developed, including NFC and RFID which are highly utilized in protocols in healthcare applications, mobile data exchange, social media, entertainment, and educational applications [24].

Transporter, reader, and antenna are the three main components of these two protocols. Readers have carried Radio Frequency (RF) signals toward tags in order to be modulated based on the antenna's impedance. The modulated signal which has been returned to the reader is analyzed to discover the exact tag [25]. RFID tags have been classified into active and passive categories based on their source of power. NFC operates at a lower frequency than RFID since it utilizes passive tags. Table 1 summarizes the specifications of NFC and RFID.

**Table 1. RFID and NFC specifications**

|  | NFC | RFID |
|---|---|---|
| Range | 5 cm | 100 m |
| Frequency band | 13.56 MHz | 3.11 GHz |
| Data rate | 424 Kbps | 4 Mbps |
| Encoding/decoding | Manchester coding | On-off keying |
| Memory size | 4 Kbytes | 32 Kbytes |

*B. Short and Medium Range Protocols*

Communication in the short-range distance (usually less than 300 meters) is covered by different network protocols, including Bluetooth, Zigbee, Z-Wave, EnOcean, Wireless Highway Addressable Remote Transducer (Wireless HART) Protocol, and Wireless Fidelity (WiFi).

Bluetooth has been classified into four classes based on the range (5 cm to 100 meters) and the transmit power (0.5 to 100 mW) based on IEEE 802.15.1 standard. It uses multiple equal channels to prevent coexisting while the frequency band varies from m 2.4 GHz to 2.485 GHz [26].

As a low-power and low-rate protocol, Zigbee has been designed based on digital radios and according to IEEE 802.15.4 standard. This protocol typically uses in Home Area Networks (HAN) and operates in the 2.4 GHz frequency band, consuming 185 microwatts per second. The nominal range and data rate of Zigbee are 100 meters and 250 Kbps, respectively [27].

Z-Wave and EnOcean are two similar low power wireless communication systems since both operate in the same range (300 meters) and frequency band (868 MHz). EnOcean is more efficient than Z-Wave because it provides a 125 Kbps data rate (85 Kbps higher than Z-Wave) while its power consumption is significantly low.

WirelessHART is the first open wireless protocol that has been developed for process measurement and control applications, based on IEEE 802.15.4 standard [28]. WirelessHART operating range is 250 meters, while its data range and frequency bands equal Zigbee.

WiFi is an extremely employed protocol based on the specifications in IEEE 802.11 standard, functioning in between 2.4 GHz and 5.6 GHz industrial, scientific, and medical (ISM) bands [29]. WiFi has the highest data rate in the short-range protocols by far (600 Mbps for ratification IEEE 802.11n). The maximum outdoor range for WiFi estimates 250 meters which is higher than Zigbee, equals WirelessHART, and less and EnOcean.

*C. Long-Range Protocols*

Sigfox, LoRaWAN, and Neul are three long-range and Low-Power Wide Area Networks (LPWAN) standards. IoT devices that employ these protocols are cheap and consume a very low amount of energy [30].

The main aim of utilizing Sigfox is to the increasing range while minimizing power consumption. It operates in a radio band of 900 MHz with a maximum range of 10 km in urban environments. The maximum data rate if Sigfox is 600 bps which is very low compared to other protocols.

LoRaWAN, similar to Sigfox, provides bidirectional communication in smart farming applications and Machine 2 Machine (M2M) interactions. LoRaWAN uses 928 MHz radio bands, delivering service in a maximum range of 15 km in a suburban environment. Although the range is less than LoRaWAN, a data rate of 50 Kbps is the superiority of this protocol. Moreover, LoRaWAN is capable of removing noise and interference due to utilizing the chirp spread spectrum method [31].

Comparable with Sigfox and LoRaWAN, Neul operates at a radio frequency band of 900 MHz, providing an average range of 10 km and a data rate of 100 Kbps. The outstanding advantage of Neul as a weightless protocol is the incredibly tiny power consumption as little as 20 mA [32].

*D. Cellular Networks*

Communication between all tools and devices across a smart farming platform can also be performed through cellular networks, including 2G, 3G, 4G, and 5G in the near future. Data range has been enhanced dramatically from 171 kbps for 2G to 2 Mbps and 100 Mbps in 3G and 4G, respectively [33]. Ultimately, 5G technology has been introduced that is capable of transmitting data up to 10 Gbps. Although the range of other protocols is almost 16 km, 5G has a limited range of fewer than 500 meters. Also, 5G operates at a radio frequency of 39 GHz, which is considerably higher than employed frequencies by other cellular technologies [34]. This technology still needs to be enhanced since the signal can be interrupted by physical obstructions like buildings, towers, and trees, just to name a few. Also, developing 5G infrastructure or upgrading existing networks required a considerable capital cost.

## III. POTENTIAL CYBER ATTACKS IN THE COMMUNICATION LAYER

Smart farming security vulnerabilities in the communication layer are similar to other IoT domains considering the structural design and protocols of the utilized technology. For illustration, a WiFi encrypted is more vulnerable to password cracking than other protocols [35]. Fig. 3 shows, security challenges in the communication layer are classified into two main categories, including data attacks and network and equipment attacks [21].

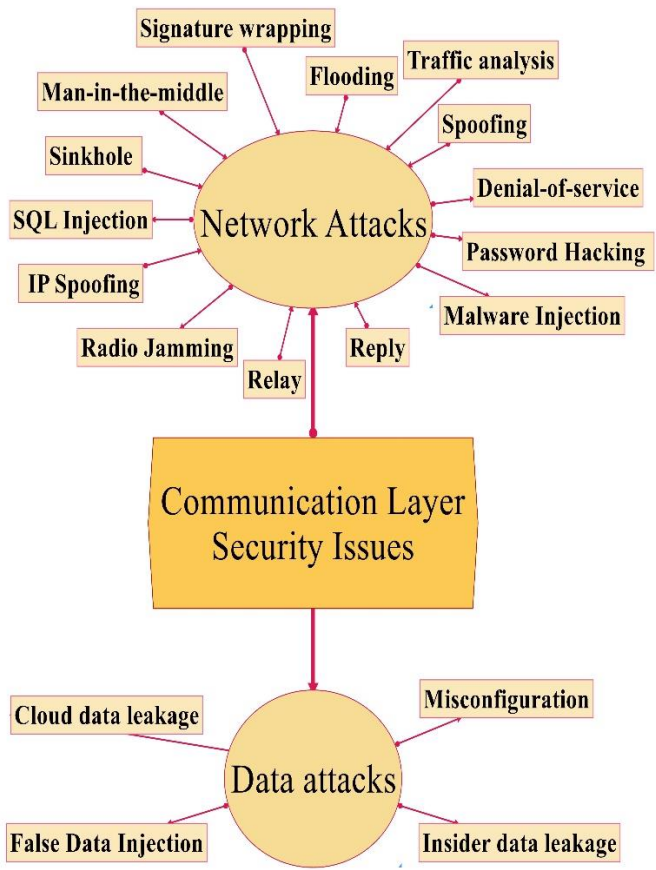

Fig. 3 Cybersecurity challenges in the communication layer

This study briefly introduces smart farming communication layer attacks as follows [36]–[41]:

- Denial-of-Service (DoS) attack: DoS happens by flooding the network with an overwhelming amount of information. DoS is the most probable attack in a smart farming framework.
- Malware injection attack: Installing harmful software that performs unauthorized actions with the aim of causing damage to the network.
- Sinkhole attack: An intruder, who is an insider attacker, tries to discover neighbor nodes routines in order to attract their traffic. This attack affects every single node in the network regardless of the distance from the base station.
- Phishing attack: An attacker fakes itself as a trusted source while trying to steal sensitive information. The victim is tricked through various methods, including clicking on a malicious link or replying to a text message, just to mention a few.
- Hello flood attack: An attacker broadcasts or replays a Hello packet with extremely high transmitting power so that most of the nodes take it as a parent node.
- Hacking attacks: By deciphering passwords, an intruder gains access to the network. There are several hacking techniques involving brute force, dictionary, rainbow table, credential stuffing, and password spraying attacks or even social engineering.
- Internet Protocol (IP) spoofing attack: An attacker creates IP packets used to trick other devices by modifying or hiding the source address.
- SQL injection attack: A web security vulnerability leads to a malicious activity via executing SQL query statements. The attacker may steal, modify, or delete the database server contents.
- Reply attack: This attack happens by eavesdropping into transmitted information in a communication network, then fraudulently sending those data to misdirect the victim.
- Man-in-the-Middle (MITM) attack: An intruder intercepts the data transmission between the communicating nodes since the nodes cannot verify each other's status.
- Relay attack: This attack is reminiscent of reply and MITM attacks where an attacker intercepts and relays information to a third party without observing or altering it.
- Radio jamming attack: In fact, this attack is a type of DoS attack in wireless networks. Radio signals are disrupted by shrinking the Signal-to-Inference-plus-Noise Ratio (SINR) using a constant jammer, deceptive jammer, random jammer, or reactive jammer.
- Traffic analysis attack: This attack does not compromise the network's data, and attackers are just capable of observing network traffic statistics or monitoring transmitted information.

*A. Ultra-Short-Range Protocols security analysis*

RFID and NFC vulnerabilities in communication levels are classified into two main classes: channel and system security risks. RFID and NFC technologies are extremely vulnerable to relay attacks since this technology uses air gap as the communication medium, which provides a spying opportunity. Also, MITM and reply attacks are other possible attacks taking advantage of channel vulnerability. System threats may lead to password hacking, DoS, and malware injections. Since encoding is a method to guarantee confidentiality and integrity in RFID and NFC, password hacking is the most common attack arising from system vulnerabilities. DoS attacks are mostly physical threats involving kill command, de-synchronization, and tad data modification attacks, but it affects the system performance as well [42]. Although current ultra-short-range technologies are not capable of installing software due to lack of memory capacity, in the near future, malware injection could be a potential attack [43].

*B. Short and Medium Range Protocols Security Analysis*

Bluetooth vulnerabilities are varied based on the utilized version, but the communication security level degrades to the weakest link. This study presents the joint vulnerabilities in all versions. Encryption length and function have a direct relationship with the risk of a hacking attack. Since the minimum required encryption length in Bluetooth is just one byte, hacking is one of the most prevalent attacks in all versions of this technology. In the absence of user authentication and limitation for device authentication, MITM and relay are two other potential attacks in Bluetooth technologies. The two last-mentioned attacks may also lead to MAC address spoofing attacks since the header of transmitted information contains MAC addresses of the devices [44].

ZigBee claims to deliver state-of-the-art security applying seven protection levels, including physical, Medium Access Control layer (MAC), network, application support, application framework, device object, and security service provider layers. However, low computing power makes ZigBee more vulnerable to network attacks. The most notorious attacks in this technology are DoS, reply, flooding, insider and jamming attacks [45]. Flooding attacks, in fact, are pre-attacks that assess the network security to perform DoS attacks. Insider attack occurs by blocking communication among devices that can manipulate the routing protocols. Reply attack simply happens in a ZigBee network where encryption is not enabled or in the lack of trust center.

Z-wave technology is vulnerable to spoofing, MITM, and sinkhole attacks. Spoofing attacks rising from the controller or another device since the source and destination of the MAC Protocol Data Unit (MPDU) frame have been known as a trusted field. A MITM attack is another threat due to the mesh network design of the Z-wave. Ultimately, exploiting the impersonation and source route cache modification vulnerabilities may lead to a sinkhole attack [46].

Although employing a unique, unalterable 32-bit identification number in EnOcean provides secure communication, this technology is still vulnerable to MITM and IP spoofing attacks. Anti MITM attack that has been proposed for EnOcean is activated just for line-powered devices. IP spoofing attack exploits mesh design and the IP weaknesses of EnOcean [47].

Improper input validation besides open physical medium makes WirelessHART a potential target of jamming attacks. Since the security of WirelessHART relies on the safety of the join key, an insider attack consequently can be another potential threat [48].

WiFi is most susceptible to password hacking attacks due to cryptographic weaknesses in Wired Equivalent Privacy (WEP), leading to a MITM attack as well. Also, taking advantage of interference that originated from a wireless environment, an attacker can combine jamming techniques in order to launch a DoS attack. Moreover, a MAC spoofing attack can be easily performed since an attacker is able to bypass filtering control. Finally, malware injection is another potential threat exploiting buffer overflow conditions [49], [50].

The 5G TCP/IP protocol has several vulnerabilities leading to DoS and MITM attacks [51]. A DoS attack in cellular networks originated from an SYN flood where the server faces numerous SYN requests, while MITM launches when an attacker captures the initial handshake among network devices compromising private communications. Lastly, a jamming attack runs an intentional interference in cellular networks via a jammer which is a malicious node. There are several types of jammers, including regular, delusive, random, responsive, control channel, and go-next jammers [52]. Table 2 presents the most important weakness of different communication protocols in a smart farming platform beside the most frequent associated cyber-attack.

**Table 2. Major weakness and potential attack based on the protocols.**

| Protocol | Weaknesses | Frequent attack |
| --- | --- | --- |
| NFC | Data exposure | Relay/MITM |
| RFID | Data exposure | Relay/MITM |
| Bluetooth | Key exchange protocol | Hacking/Relay |
| ZigBee | Access Control | DoS/Flooding |
| Z-wave | Certificate Validation | MITM/Sinkhole |
| EnOcean | Identification | IP spoofing |
| WirelessHART | Input validation | Insider |
| WiFi | Cryptography | Password Hacking |
| LoRaWAN | Input validation | DoS/Jamming |
| Sigfox | NAK key | Reply |
| Neul | Input validation | Traffic analysis |
| Cellular | Input validation | Jamming/DoS |

## IV. CONCLUSION

Food security is a major concern and global challenge due to fast population growth along with global warming. Accordingly, utilizing new technologies and smart equipment is inevitable. Taking advantage of IoT, intelligent farming frameworks provide local and global connectivity among all devices in order to increase yield and quality of products through real-time monitoring, controlling irrigation, recording moisture and temperature, and adapting treatments to these and other monitored inputs. However, high penetration of smart devices along with bidirectional communication leads to novel security concerns.

In this review paper, various utilized communication technology in smart farming has been classified into four main classes: ultra-short-range, short and medium-range, long-range, and cellular networks. All the technologies have been investigated considering standard protocol, radio frequency band, range, and data rate to discover advantages and disadvantages. Potential attacks in the communication layer were presented. Finally, the most frequent attacks for specific vulnerabilities in four different communication technology groups were identified.